\documentclass[longnamesfirst,apjl]{emulateapj}
\usepackage{apjfonts,natbib,epsfig}
\tighten

\newcommand{\beq}{\begin{equation}}
\newcommand{\eeq}{\end{equation}}
\newcommand{\bea}{\begin{eqnarray}}
\newcommand{\eea}{\end{eqnarray}}

\begin{document}
\title{Steady-State Electrostatic Layers from Weibel Instability in Relativistic \\ Collisionless Shocks}
\author{Milo\v s Milosavljevi\'c$^1$, Ehud Nakar$^1$, 
and Anatoly Spitkovsky$^{2,3}$}
\affil{$^1$Theoretical Astrophysics, California Institute of Technology, Pasadena, CA 91125 \\ 
$^2$ Kavli Institute for Particle Astrophysics and Cosmology, Stanford University, P.O.\ Box 20450, MS 29, Stanford, CA 94309 \\ 
$^3$ Chandra Fellow }
\righthead{COLLISIONLESS SHOCKS}
\lefthead{MILOSAVLJEVI\'C, NAKAR, \& SPITKOVSKY}

\begin{abstract}

It is generally accepted that magnetic fields generated in the nonlinear development of the transverse Weibel instability provide effective collisionality in unmagnetized collisionless shocks.  Recently, extensive two and three dimensional simulations improved our understanding of the growth and saturation of the instability in colliding plasma shells.  However, the steady-state structure of the shock wave transition layers remains poorly understood.  We use basic physical considerations and order-of-magnitude arguments to study the steady state structure in relativistic unmagnetized collisionless shocks in pair plasmas. The shock contains an electrostatic layer resulting from the formation of stationary, magnetically-focused current filaments. The filaments form where the cold upstream plasma and the counterstreaming thermal plasma interpenetrate. The filaments are not entirely neutral and strong electrostatic fields are present. Most of the downstream particles cannot cross this layer into the upstream because they are trapped by the electrostatic field.  We identify the critical location in the shock transition layer where the electromagnetic field ceases to be static.  At this location, the degree of charge separation in the filaments reaches a maximum value, the current inside the filaments comes close to the Alfv\'en limit, and the phase space distribution function starts to isotropize. We argue that the radius of the current filaments upstream of the critical location is about twice the upstream plasma skin depth.  Finally, we show that some downstream particles cross the electrostatic layer and run ahead of the shock into the preshock medium without causing instability.  These particles may play an important role in particle acceleration.


\end{abstract}

\section{Introduction}
\label{sec:introduction}

Relativistic collisionless shocks should accompany all
relativistic outflows, such as those associated with
active galactic nuclei (AGN), microquasars, gamma-ray bursts (GRBs),
giant flares from soft gamma-ray repeaters, and pulsar wind
nebulae. Nonrelativistic collisionless shocks are also believed to
be present in cosmological structure formation, supernova remnants,
and planetary bow shocks. The composition of the
relativistic outflow (e.g., $e^-e^+$, $e^-p$, or Poynting flux) is
not securely determined in any of the relativistic sources, and
the level of magnetization is poorly constrained.
In some systems (e.g., AGN) it seems that the
magnetic fields in the upstream medium of the shock are dynamically
important, while in others (e.g., ``external'' GRB shocks), the
upstream is effectively unmagnetized. In the latter case, 
the magnetic fields that are
invoked to explain the observed synchrotron emission seem to be
generated in the shock itself. 

Magnetic fields with energy density
near equipartition with the kinetic energy density of the plasma can
be generated by the nonlinear development of the transverse Weibel
instability.  The Weibel instability \citep{Weibel:59} arises 
when the velocity distribution of
the plasma particles is anisotropic (e.g., \citealt{Yoon:87}).  
The transverse Weibel instability (e.g., \citealt{Fried:59,Silva:02,Bret:04,Bret:05})
results in the breakdown of the plasma 
into current filaments \citep{Lee:73}.  Strong
anisotropy arises in collisions in which charged particles
interpenetrate. This has naturally led to suggestions that the transverse
Weibel instability is responsible for the growth of magnetic fields
in unmagnetized collisionless shocks
\citep{Medvedev:99,Gruzinov:01}.  In this picture, magnetic fields
generated in the shocks provide effective collisionality in the
otherwise collisionless transitional layer of the shock. The
magnetic fields may also be responsible for diffusive particle
acceleration in the shock.

The growth and saturation of the transverse Weibel instability in colliding cold unmagnetized plasma shells (both $e^-e^+$ and $e^-p$) has been
extensively studied using three-dimensional particle-in-cell (PIC) simulations
(\citealt{Silva:03,Fonseca:03,Frederiksen:04,Jaroschek:04,Jaroschek:05,Nishikawa:05}; Spitkovsky \& Arons 2005, in preparation). These simulations are
computationally demanding; they must resolve multiple plasma skin
depths in the direction perpendicular to shock propagation (henceforth, perpendicular direction), and a very large
number of plasma skin depths in the direction 
parallel to shock propagation (henceforth, parallel direction)
over which field growth and plasma isotropization and thermalization
take place. PIC simulations have not yet converged to reveal the
steady-state structure of the transition layer that persists after
the transients associated with the initial collision of plasma
shells have dissipated.

Simulations of cold shell collisions show that current
filaments form parallel to the axis of collision
(e.g.,~\citealt{Silva:03,Frederiksen:04}). The direction of current flow in
neighboring filaments alternates. The currents are maintained by
the pinching force of the self-magnetic field.  The maximum
current that can flow through any given filament, precluding
unphysical velocity-space configuration, is the Alfv\'en limiting
current \citep{Alfven:39,Davidson:74,Honda:00}.  The 
development of the transverse Weibel instability in cold, colliding shells
involves the merging of current filaments \citep{Medvedev:05} and
saturation of the Alfv\'en limit \citep{Kato:05}.

It has long been known that it is possible to construct exact,
albeit unstable, solutions for collisionless shock transition layers
in one dimension, where kinetic energy of the upstream plasma is
converted into electrostatic potential energy (e.g., of plasma
oscillations) as the plasma crosses the transition layer
\citep{Moiseev:63,Montgomery:69}. These solutions invoke particle reflection by the stationary electrostatic
potential of the transition layer and belong to
the class of ``double layers.'' A double layer is an electrostatic structure which can appear within a current carrying plasma and sustain a significant net potential difference associated with at least two layers of opposite net charge (\citealt{Raadu:89} and references
therein).  

\citet{Moiseev:63} noticed that
the interpenetration of the reflected and the forward streaming
plasma is unstable to a Weibel-like mode.  They conjectured that the
scattering of particles by the electromagnetic fields generated in
the instability gives rise to an effective viscosity that
facilitates momentum diffusion in the transition layer.  Albeit
prescient, this model has not developed into a self-consistent picture
of the shock transition layer.

Here we suggest a self-consistent steady state structure of a
relativistic shock that propagates into unmagnetized pair plasma. In
\S~\ref{sec:Outline}, we sketch out a qualitative physical picture of
the shock transition layer.  In \S~\ref{sec:notation}, we introduce the reference frame and notation that is used in the rest of the paper. In \S~\ref{sec:ballistics}, we discuss the motion
of particles in the quasistatic electromagnetic field of current
filaments that form in the transition layer.
In \S~\ref{sec:separation}, we study the mechanics of the separation
of the upstream flow into filaments of oppositely charged particles.
In \S~\ref{sec:counterstream}, we study the kinematics of
particles trapped by the electrostatic field in the transition
layer, and show how these particles facilitate the onset of charge
separation.   In \S~\ref{sec:diameter}, we argue that the radius of
the current filaments must  be about equal to the plasma skin depth.  In \S~\ref{sec:precursor}, we show that particles in
the high-energy tail of the thermalized downstream can escape into
the upstream domain.
In \S~\ref{sec:discussion}, we discuss the
compression and convergence to the hydrodynamic jump
conditions, and we also 
speculate on the applicability of our model to $e^-p$ shocks. Finally, in \S~\ref{sec:conclusions}, we summarize our main conclusions.

\section{Physical Picture}
\label{sec:Outline}

We consider an ultrarelativistic shock wave
propagating into unmagnetized pair plasma. The
structure of the shock transition layer is steady
in the rest frame of the shock.
In this frame the cold,
upstream plasma arrives on one side of the transition layer, and
thermalized downstream plasma leaves on the other side.  
Since the plasma is collisionless by assumption, electromagnetic fields inside the transition layer deflect, scatter, and redistribute the kinetic energies of the particles
arriving from the upstream. Steady
currents are present to support a
steady electromagnetic field.

The saturated state of the transverse 
Weibel instability, as seen in the
simulations,  consists of magnetically self-pinched current
filaments. Initial filaments are produced at the
point where the upstream encounters the transition layer. These
filaments merge to form structure on ever larger
scales. We here focus on the first generation of filaments
only, because it is key to understanding 
the subsequent evolution of the shocked plasma.
The filaments form as the magnetic field displaces the cold upstream
particles transversely, 
perpendicular to the direction of shock propagation (see, e.g., Fig.~1 in \citealt{Medvedev:99}).  
The filaments are in place when 
the displacement
becomes similar to the separation between filaments.
The filaments live quasi-statically for some distance, and then become time dependent, as the electromagnetic fields turn from quasi-static to dynamic.

The transverse 
Weibel instability requires interpenetrating, mutually counterstreaming
plasma components.
One of the
components is the cold, upstream plasma. The other component
is the downstream plasma, i.e., 
the particles originating in the upstream that are reflected in
the downstream or in the transition layer 
and are now moving backward, toward the upstream 
in the shock frame.  Hereafter we
refer to the latter particles as the ``counterstream.'' The transition layer is where the upstream and the counterstream interpenetrate.

The small-scale toroidal self-magnetic field of the current filaments cannot keep the
counterstream particles trapped in the transition layer.  This is because the field configuration is open-ended facing the upstream and any particles affected by the magnetic field alone
would escape by bouncing between the magnetic walls until they have reached the opening. However, most of the counterstreaming particles 
have to be prevented from escaping into the upstream if hydrodynamic jump conditions are to be
reproduced.
We show that a parallel 
electrostatic field is present 
in the first generation of current filaments and 
that this field can
contain the counterstream.

Since the filaments are quasi-static, the electric field can be
derived from a potential associated with a quasi-static distribution of charge. 
The sign of the charge in neighboring filaments alternates, 
ensuring overall charge neutrality. Parallel
electric field arises from parallel charge gradients
within the filaments. The density
of counterstreaming particles within the filament is less
than that of the oppositely charged forward streaming particles,
implying non-neutrality.  Net charge does not prevent filamentation because only a
small amount of counterstreaming particles is sufficient to ensure that magnetic pinching can overcome electrostatic repulsion and separate the oncoming upstream particles  based on their charge.
We show that the degree of charge-separation reaches maximum  at about the same point where the current reaches the Alfv\'en limit.   In a self-consistent picture, the radius of the first-generation filaments is about equal to the plasma skin depth of the upstream. 



We refer to the region upstream of the point of maximum charge separation as
the ``charge separation layer.''   In this paper we investigate the
structure and properties of the charge separation layer in
detail. The layer is located at the leading edge of the shock 
before most of the density compression. 
We present a picture in which the charge separation layer is
described by an electrostatic double-layer 
that forms inside each magnetically-pinched current filament. The layer therefore functions as a diode protecting the upstream from the downstream. The filaments
themselves are a stationary end-product of the nonlinear transverse Weibel
instability. 

\begin{figure}
\plotone{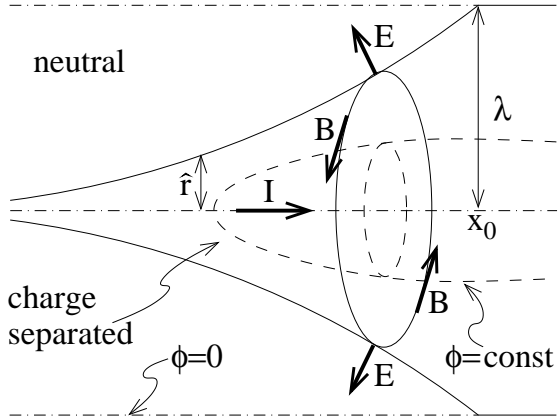}
\caption{Schematic representation of a positron within the charge separation layer; an electron filament would equivalent with the direction of the magnetic field ${\bf B}$, electric field ${\bf E}$, and current ${\bf I}$, reversed. The conical envelope ({\it solid lines}) is the trajectory of the last electron to leave the growing positively-charged domain ({\it solid ellipse}). The top and bottom dot-dashed lines separate the positron filament from the neighboring electron filaments.  The central dot-dashed line is the filament center. We also show an equipotential surface ({\it dashed lines}). \label{fig:drawing}}
\end{figure}

Figure \ref{fig:drawing} schematically depicts the structure
of a current filament in  the charge separation layer. The
upstream particles are separated into filaments with a typical radius $\lambda$
of the order of the upstream plasma skin depth. We set the electrostatic potential
to  zero at $x=-\infty$ (far left in the figure). The potential is also zero at the midpoint between filaments (the top and bottom dot-dashed line in the figure). A filament of positive current, carried
mainly by the positrons from the upstream, produces a potential ridge (central dot-dashed line), while a
filament of negative current produces a potential valley. As the magnetic field displaces an upstream 
positron  toward the positive filament, the positron
climbs the electrostatic ridge. 
 An electron located in what is to become a positron filament is displaced out (solid line in the figure).

The absolute height 
of the electrostatic potential at the point of maximum charge
separation ($x_0$ in Fig.~\ref{fig:drawing}) is comparable to the kinetic energy of
an upstream particle. This is required for the counterstream to be decelerated and reversed, as the kinetic energy of the downstream particles, among which the counterstream particles originate, is smaller by a factor of $\sim3\sqrt{2}$ than that of the upstream particles. 
The same potential decelerates the upstream particles as they cross the
charge separation layer (see Fig.~\ref{fig:upstream_trajectory}). If the upstream flow remains relativistic as it traverses the layer, compression must occur in the downstream, where time-dependent electromagnetic fields can scatter and isotropize the particles. 

To be contained within the transition layer of the shock, 
most of the counterstream must already be charge-separated as it enters 
the charge separation layer at $x_0$.  
Namely, a counterstream positron must be contained within the potential valley 
of the filament carrying upstream electrons, and vice versa (an equipotential surface is outlined with dashed lines in Fig.~\ref{fig:drawing}). 
As the positron attempts to climb out of the potential valley of the electron filament, 
it reverses
direction and proceeds back toward the downstream (see Fig.~\ref{fig:downstream_trajectory}). 

If the energy distribution of the  counterstream particles
is thermal at $x_0$, a small
fraction of the particles in the exponential tail have sufficient energy to escape the transition layer and run
ahead of the shock. We show that 
plasma stability tightly limits
the flux of counterstreaming particles that can run ahead of the
transition layer into the untouched upstream.  These particles may play important role in
diffusive particle acceleration in shock waves.

\section{Notation}
\label{sec:notation}

We work in the reference frame in which the shock is at rest.  Unless otherwise stated, all densities, velocities, momenta, and fields are written in this frame. It is common in the literature, however, to write particle densities in the local rest frame of the plasma (``fluid frame''), and to write momenta in the rest frame of the upstream (e.g., \citealt{Blandford:76}).  Lorentz factor of the unshocked upstream medium in the rest frame of the shock, which is usually denoted by $\Gamma$, we denote with $\gamma_{-\infty}$. Similarly, we denote the Lorentz factor of the downstream with $\gamma_{+\infty}$. The densities of the upstream and downstream medium far from the transition layer are denoted with $n_{-\infty}$ and $n_{+\infty}$, respectively.  

In Table \ref{tab:notation}, we express these quantities in terms of the notation of \citet{Blandford:76}, where $\gamma_2$ is the Lorentz factor of the downstream in the rest frame of the upstream; $n_1$ and $n_2$ are the rest frame densities of the upstream and the downstream, respectively, and $\hat\gamma_2$ is the ratio of specific heats.  Note that hydrodynamical jump conditions dictate that $n_{+\infty}/n_{-\infty}=3$. All quantities referring to the counterstream particles are marked with a prime.  Let $x_0$ denote the point of maximum charge separation; the quantities measured at $x_0$ are marked with subscript zero. 

\begin{deluxetable}{lcc}
\tablecolumns{3}
\tablecaption{Notation \label{tab:notation}}
\tablehead{
\colhead{Quantity} &
\colhead{Our Notation} & 
\colhead{Ultrarelativistic Limit}
}
\startdata
upstream in shock frame & $\gamma_{-\infty}$ & $\sqrt{2}\gamma_2$ \\
downstream in s.\ f.\  & $\gamma_{+\infty}$ & $3\sqrt{2}/4$ \\
upstream density in s.\ f.\ & $n_{-\infty}$  & $\sqrt{2}\gamma_2n_1$ \\
downstream density in s.\ f.\ & $n_{+\infty}$  & $3\sqrt{2}\gamma_2n_1$ \\
compression factor in s.\ f.\ & $n_{+\infty}/n_{-\infty}$ & $3$ \\
\enddata
\end{deluxetable}

\section{Ballistics}
\label{sec:ballistics}

We here consider the ballistics of a particle moving inside a current filament.
We use cylindrical coordinates $(x,r,\theta)$ where
$x$ is the symmetry axis of the filament 
parallel to the direction of shock propagation.
Since counterstream particles have large perpendicular velocities ($\beta_r\sim 1$),
they cross the middle of the filament many times.  Therefore,
the quantities pertaining to the counterstream particles can be evaluated at 
the center of the filament $r=0$, e.g., $\beta'(x)\equiv
\beta'(x,0)$, where the point of maximum charge separation $x_0$ is where an average particle has executed one oscillation toward the center of a filament. 
The upstream particles on average cross 
the filament only once between $x=-\infty$ and $x_0$ 
and thus their radial dependence must be retained.

We approximate a filament with a
cylinder of radius $\lambda$.  The filament touches 
neighboring filaments of the
same radius that carry equal and opposite current. In reality, a filament interfaces with the neighboring filaments over its entire area in such a way that no  empty space remains between. Cylindrical filaments are thus an idealization.
We assume that the properties of the filament (net current, charge
density, etc.) vary along the $x$ axis on scales much larger than
$\lambda$, implying that the filament is locally two-dimensional.

Because of the quasi-static nature of the filaments, the total energy
of a particle, $\gamma m c^2+q\phi$, is conserved. Here, $q$ is the
charge of the particle, $\gamma$ is the Lorentz factor, $m$ is the
mass of the particle, and $\phi$ is the electrostatic potential.  
Conservation of energy for a particle originating in the upstream then implies
\begin{equation}
\label{eq:energy_conservation}
\gamma_{-\infty} mc^2 = \gamma(t)mc^2 + q\phi(x(t),r(t)) .
\end{equation}
A cold upstream particle is deflected slowly towards the center of
its filament. The equation of motion for the perpendicular momentum of
the particle reads
\begin{equation}
\frac{dp_r}{dt}=q(E_r- \beta_x B_\theta) ,
\end{equation}
which, using $p_r=\gamma m c dr/dt$ and $d/dt=c\beta_x d/dx$, can be
rewritten as
\begin{equation}\label{eq:separation_derivx}
m c^2 \beta_x \frac{d}{dx}\left(\gamma \beta_x \frac{dr}{dx}\right)=
q(E_r- \beta_x B_\theta) .
\end{equation}
Note that for the current to pinch, the condition $\beta_x B_\theta>E_r$
must be satisfied, implying that there exists 
a local reference frame in which the filament is charge-neutral and $E_r$ vanishes.
We call this frame ``the frame of the current.''  Let the velocity relating the frame of the shock to the frame of the current be $\tilde\beta=-E_r/B_\theta$, and the corresponding Lorentz factor $\tilde\gamma$. 

A counterstream particle is electrostatically 
confined to the filament if 
$(\gamma_0'-1)mc^2<-q'\phi_0$, 
where $\gamma'_0$ is the Lorentz factor of the counterstream
particle at $(x,r)=(x_0,0)$ and $\phi_0$ is the potential at this
location. A particle that is electrostatically confined to its filament
is also bound to the transition layer and it must decelerate and reverse
its direction (see Fig.~\ref{fig:downstream_trajectory}). A
particle which is not confined electrostatically may still be
confined magnetically to the filament and reversed
if $\beta'_{x0} \ll \beta'_{r0}$.  A counterstream particle 
that is not confined to the filament runs away to $x=-\infty$.

Near $x_0$, charge separation is almost complete and the electromagnetic field varies weakly with radius.  We consider an upstream particle that is moving ultrarelativistically inside a filament in the same sense as the current in the filament, i.e., $\beta\tilde\beta>0$ and $\gamma\gg\tilde\gamma\gg1$.  The magnetic field tends to confine the particle within the current filament, while the electric field tends to repel it out of the filament.  Given that $|E_r|<|B_\theta|$, the magnetic field wins by a small margin, and confinement is possible provided that the particle's momentum is not too large.  It is straightforward to show that the particle remains confined provided that $\beta_r^2 \lesssim \lambda eB_\theta/\gamma\tilde\gamma^2 mc^2$.

\section{Charge Separation}
\label{sec:separation}

We here derive an equation governing the degree to which the charges in the upstream are consigned to disjoint current filaments.  Since the upstream plasma is neutral in the limit $x\rightarrow-\infty$, the region $r<\lambda$ in which the pinching of charges $q$ takes place must be evacuated of charges $-q$.  Here, $\lambda$ is the radius of the filament.   Charge separation takes place gradually.  We trace the orbit $(\hat x(t),\hat r(t))$ of the last of the charges $\hat q=-q$ to leave the region. This particle had moved just along the $x$-axis before approaching the charge separation layer (see Fig.~\ref{fig:drawing}).  The region $r<\hat r(t)$ can then be considered ``charge-separated,'' while the region $r<\hat r(t)<\lambda$ is still neutral. 

We make the simplifying assumption that the particle density $n$ and mean axial velocity $\beta_x$ of particles in the upstream is independent of $r$.  The upstream plasma at $x=-\infty$ is undisturbed, neutral, and cold, with density $n_{-\infty}$ moving toward the shock with the Lorentz factor $\gamma_{-\infty}$ in the frame of the shock.  In this section we assume that the electromagnetic field is static, and thus that the electric field is irrotational and derived from a potential $\phi(x,r)$ which vanishes at $x=-\infty$.

The electromagnetic field of neighboring filaments cancels. Therefore the dominant contribution to the electromagnetic force comes from a segment of length $\Delta x\sim 2\lambda$ of the closest (host) filament. Since the region $r<\hat r(t)$ charge-separated, the electric field due to the upstream particles at $r=\hat r(t)$ is $E_r\sim 2\pi qn\hat r$ and the magnetic field is $B_\theta\sim 2\pi\beta_x qn r$.  Let $\mu'(x)$ and $j'(x)$ denote, respectively, the linear particle number and the particle current density of the counterstreaming particles,
\beq
\mu'\equiv \int_0^\lambda n' 2\pi rdr, \ \ \ \ j'\equiv \int_0^\lambda c\beta_x' n' 2\pi rdr .
\eeq
Then the electromagnetic field of the counterstream is given by $E_r'\sim 2q'\mu'/\lambda$ and $B_\theta'\sim 2q'j'/c\lambda$.  

Parameterizing $\hat r(t)$ in terms of $x=\hat x(t)$ and substituting the values of the fields in equation (\ref{eq:separation_derivx})  and dividing by $mc^2$ we obtain
\beq
\hat \beta_x \frac{d}{dx}\left(\hat \gamma \hat \beta_x \frac{d\hat r}{dx}\right)=
\frac{q}{mc^2} \left[-2\pi (1-\beta_x)n\hat r q- 2\left(\mu'-\frac{j'}{c}\right) \frac{q'}{\hat r}\right] .
\eeq
The initial condition is $\hat r(-\infty)=0$ and we are interested in the evolution of $\hat r(x)$ up to the point where $r=\lambda$.

Particle number conservation for upstream particles implies that $\beta_{-\infty}n_{-\infty}= \beta_x n$. Also note that $qq'=-e^2$.  Let $\bar\beta_x'\equiv j'/c\mu'<0$ denote the average velocity of the counterstreaming particles.  Then the upstream particles experience
\beq
\label{eq:separation_ninfty}
\hat \beta_x \frac{d}{dx}\left( \hat \beta_x\frac{\hat \gamma}{\gamma_{-\infty}} \frac{d\hat r}{dx}\right)=
\frac{1}{2\delta^2} \left[- \frac{(1-\beta_x)}{\beta_x}\hat r+\frac{(1-\bar\beta_x') \mu' }{\pi n_{-\infty}\hat r}\right] ,
\eeq
where we have defined the plasma skin depth of the upstream particles $\delta\equiv c/(4\pi n_{-\infty}e^2/\gamma_{-\infty} m)^{1/2}$ and have divided the equation by $\gamma_{-\infty}$. 

Note that $\phi(\hat x,\hat r)$ vanishes as $\hat r$ approaches $\lambda$.  Therefore it is a reasonable approximation to set the potential at the boundary of the charge-separated region to zero, which implies that $\hat\gamma\sim \gamma_{-\infty}$ is a constant and the upstream remains ultrarelativistic, $\hat\beta_x\sim\beta_{-\infty}\sim 1$.  Let $X\equiv x/\delta$, $R\equiv\hat r/\delta$, $L\equiv \lambda/\delta$, and $N'\equiv\mu'/\pi\delta^2n_{-\infty}$; note that the latter quantity is equal to $\gamma_{-\infty}/4$ times the so-called Budker's parameter.  Then equation (\ref{eq:separation_ninfty}) can be rewritten as
\beq
\label{eq:separation_variables}
\frac{d^2R}{dX^2} =
-\frac{ (1-\beta_x)}{2\beta_x} R + \frac{(1-\bar\beta_x') N' }{2  L} ,
\eeq
with the initial value $R=0$ and final value $R\sim L$. 

Equation (\ref{eq:separation_variables}) describes the mechanics of charge separation among the upstream particles. The charge separation, and thus the self-pinching of the current filament, is possible only if the right hand side of the equation is positive, or
\beq
N'> \frac{(1-\beta_x)}{(1-\bar\beta_x')\beta_x} RL .
\eeq  
Evidently, without the presence of a small amount of counterstreaming particles, the current does not pinch, and the filament does not form.
Equation (\ref{eq:separation_variables}) can be solved once the density of counterstreaming particles, $N'$, is self-consistently specified.

\begin{figure}
\epsscale{1.2}
\plotone{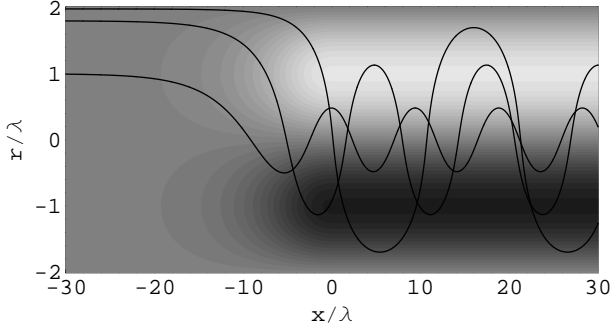}
\caption{Trajectory of three representative upstream particles in the electromagnetic field of the toy model (see text). Initial radial displacements of the particles are $r(-\infty)=(1,1.8,1.98)\lambda$. The grayscale shows the magnetic field, which vanishes at $x=-\infty$ and $r=0$.  The radial coordinate was extended to negative values to cover the meridional plane.\label{fig:upstream_trajectory}}
\end{figure}

We use a toy model to integrate the trajectory of an upstream particle.  The electrostatic potential and the magnetic field are taken to be
\bea
\label{eq:toy_model}
\phi(x,r)&=&C\gamma_{-\infty}\tilde\beta\left[\frac{4e^{x/\Delta}}{(1+e^{x/\Delta})^2} H(-x) + H(x)\right]\cos\left(\frac{\pi r}{2\lambda}\right) , \nonumber\\
B_z(x,r)&=& \frac{C\gamma_{-\infty}\pi}{2\lambda}\left[\frac{4e^{x/\Delta}}{(1+e^{x/\Delta})^2} H(-x) + H(x)\right] \sin\left(\frac{\pi r}{2\lambda}\right),
\eea
with $\Delta=4\lambda$, $\tilde\beta=0.9$, $\gamma_{-\infty}=20$, and $C=0.85$ is a factor taking into account that final charge separation is not complete.  Representative trajectories are shown in Figure \ref{fig:upstream_trajectory}.  Note that the particle with initial axial distance $r(-\infty)=1.98\lambda$ crosses $r=\lambda$ near $x=0$, where charge separation must be maximum. 

\section{The Counterstream}
\label{sec:counterstream}

The phase space distribution of the counterstream particles depends on the dynamics of the downstream $x>x_0$ where the electrostatic approximation breaks down.  Therefore, the distribution cannot be derived self-consistently within the scope of the present work.  Instead, one can attempt to guess the density of the counterstream particles as a function of position.  The most natural choice is that in which the density $\mu'$ is a function of the electrostatic potential only, because $\phi(x)$ determines how many counterstream particles are energetic enough to reach $x$. There is not a unique distribution giving rise to a specific potential. The simplest density profile motivated by Maxwellian-like distributions is $\mu(\phi)\propto e^{-q'\phi/\gamma_0mc^2}$, where $\gamma_0$ is a constant.

Since $\phi(-\infty)=0$, an exponential density profile implies that a finite number of unbound counterstreaming particles exist in the upstream.  We here select only the particles that are confined by the electrostatic potential.  We will return to the particles belonging to the high-energy tail of the Maxwellian, which are not confined, in \S~\ref{sec:precursor}.  Some particles can also be magnetically confined to a part of the filament even if they are not electrostatically confined.  The particles that are not confined are not charge separated into filaments. 
Linear density of counterstreaming particles is given by
\beq
\label{eq:mu}
\mu'(x) = \mu'(x_0)
\frac{e^{-q'\phi(x)/\gamma_0 m c^2}-1}{e^{-q'\phi(x_0)/\gamma_0 m c^2}-1} ,
\eeq
where $\phi(x)$ is the electrostatic potential at $(x,r=0)$, which depends on 
the charge density via
\bea
\label{eq:potential}
\frac{q'\phi}{mc^2}&\sim& \frac{q'}{mc^2} \{\pi \hat r^2 [2\ln(\lambda/\hat r)+1] n q + 2\mu' q'\}\nonumber\\
&= &\frac{\gamma_{-\infty}}{2} \left\{-  \frac{R^2}{\beta_x}\left[\ln\left(\frac{L}{R}\right)+\frac{1}{2} \right] +  N'\right\} .
\eea
The  density of the counterstreaming particles and the electrostatic potential are transcendental solutions of equations 
(\ref{eq:mu}) and (\ref{eq:potential}).  

The situation simplifies greatly if $N'\ll L^2$ and we can ignore the contribution of the counterstreaming particles to the potential.  Of course, we must retain the contribution of the counterstreaming particles to the magnetic field in  equation (\ref{eq:separation_variables}).  To simplify this equation, note that deep in the upstream $R\ll L$, while $\beta_x\sim 1$ and $(1-\beta_x)\sim 1/2\gamma_{-\infty}^2$. Since $1<(1-\bar\beta_x')<2$, to an order of magnitude we can replace this quantity with $2$.  Thus, after substituting equations (\ref{eq:mu}) and (\ref{eq:potential}) in equation (\ref{eq:separation_variables}) and keeping the lowest-order terms in $R$ we obtain
\bea
\label{eq:pinch_small_r}
\frac{d^2R}{dX^2} &=&-\frac{R}{4\gamma_{-\infty}^2} +  N'(x_0) \frac{\gamma_{-\infty}}{2\gamma_0} \frac{R^2}{L} \ln\left(\frac{L}{R}\right) e^{-\gamma_{-\infty}L^2/4\gamma_0}.
\eea
Current pinching is possible only when the right hand side of equation (\ref{eq:pinch_small_r}) is positive---a condition that is manifestly not satisfied for very small $R$, roughly when $R\ll 1/\gamma_{-\infty}^2$.  Roots of this apparent paradox lie in the approximate treatment of the counterstreaming particles. We believe that the paradox would be eliminated by correctly treating the full three-dimensional orbits of the particles.  

In the regime in which the second term on the right hand side of equation (\ref{eq:pinch_small_r}) is much larger than the first term, we estimate the width of the charge separation layer $\Delta$ by replacing the second derivative with a ratio, $d^2R/dX^2\sim R/\Delta^2$. If $\gamma_0$ is interpreted as the temperature of the downstream medium, conservation laws imply that $\gamma_{-\infty}/\gamma_0\sim3\sqrt{2}$.  The width of the charge separation layer in units of the plasma skin depth then reads
\beq
\label{eq:layer_depth}
\Delta \sim \frac{0.7}{\sqrt{N'(x_0)}} \left[\frac{R}{L}\ln\left(\frac{L}{R}\right)\right]^{-1/2} e^{0.53\ L^2} .
\eeq

Note that the number of counterstream particles that are {\it not} confined by the electrostatic potential is given by $N'_{\rm free}\sim N'(x_0) e^{-\gamma_{\infty}L^2/4\gamma_0}$. Therefore the square of the exponential factor in equation (\ref{eq:layer_depth}) must be much greater than unity to prevent mass exodus of the counterstream particles into the upstream. 
We will see in \S~\ref{sec:diameter}, however, that $L$ itself cannot be much larger than unity or the Alfv\'en limit would be violated. In \S~\ref{sec:precursor} we will find that stability requires $N'(x_0)\ll 1$ in shocks with very large Lorentz factors. Therefore, the transition layer is many plasma skin depths wide.  This underscores the challenges that must be overcome to reproduce the steady-state transition layers in PIC simulations, which  must resolve very many skin depths in the parallel direction.

\begin{figure}
\epsscale{1.2}
\plotone{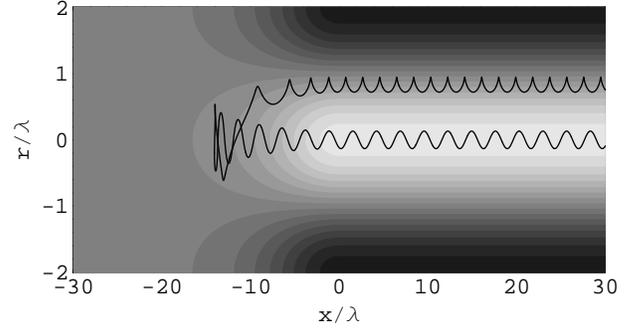}
\caption{Trajectory of a counterstream particle in the electric field of the toy model (see text). The grayscale shows the electrostatic potential, which vanishes at $x=-\infty$ and $r=\pm\lambda$. The particle has initial Lorentz factor $\gamma(+\infty)\approx15$ and initial perpendicular velocity when crossing $r=0$ of $\beta_r(-\infty)=0.3$. The radial coordinate was extended to negative values to cover the meridional plane. \label{fig:downstream_trajectory}}
\end{figure}

In Figure \ref{fig:downstream_trajectory}, we plot the trajectory of a counterstream particle in the toy-model electromagnetic field of equation (\ref{eq:toy_model}).  The particle has initial Lorentz factor $\gamma(+\infty)\approx15$, just sufficient to keep it electrostatically confined.  The initial perpendicular velocity of the particle when it crosses $r=0$ is $\beta_r(-\infty)=0.3$.  Note that the particle reverses at $x\sim-15\lambda$. 

\section{Filament Size}
\label{sec:diameter}

We have so far assumed that the radius $\lambda$ of the first generation of filaments is constant throughout the charge separation layer.  In the standard picture motivated by cold shell collisions, filaments first form with radii on scales of the fastest growing mode of the transverse Weibel instability.  The filaments immediately proceed to merge with each other in hierarchical fashion. 

In the collision of two cold shells of comparable density, the
dispersion relation for the filamentation mode derived from equation (\ref{eq:dispersion_similar_density}) by taking the limit $\beta_\perp'\rightarrow0$ reads
\beq
\label{eq:fried_relativistic}
\omega^4-c^2k^2(\omega^2+\omega_{\rm p}^2/\gamma)=0 ,
\eeq 
where $\omega_{\rm p}\equiv (4\pi e^2 n/m_e)^{1/2}$ is the plasma frequency written in terms of the total particle density in the shock frame $n$.  This is an ultrarelativistic analogue of the nonrelativistic relation derived by \citet{Fried:59}.

Equation (\ref{eq:fried_relativistic}) implies that there is a growing mode associated with every wavelength.  The growth rate decays on large scales ${\rm Im}(\omega)\propto k^{1/2}$, and saturates at the constant value ${\rm Im}(\omega)\sim \omega_{\rm p}/\gamma^{1/2}$ in the limit $k\rightarrow\infty$.  Although, formally, the infinitesimally small scales grow the fastest, the growth rate is nearly uniform on all scales larger than the plasma skin depth, i.e., for $k\gtrsim2\pi/\delta$ where $\delta\equiv \gamma^{1/2}c/\omega_{\rm p}$.  Therefore, filaments with radius $\lambda \gtrsim\delta/4$ form directly in the linear development of the transverse Weibel instability in cold shell collisions. 

A powerful argument, referring to the steady-state structure of the transition layer, argues for filamentation at the scale of the plasma skin depth.  We showed in \S~\ref{sec:counterstream} that the non-neutrality of the current filaments gives rise to an electrostatic potential that keeps most of the thermalized particles from escaping into the upstream domain.  For the confinement to be possible, the depth of the potential well $|\phi| \sim \pi \lambda^2 \chi n e$ must be larger than the temperature of the downstream medium $\sim\gamma_0 mc^2$, where $\chi\lesssim 1$ is the fraction of particles confined to a filament when charge separation is maximum and $\gamma_0\sim\gamma_{-\infty}/3\sqrt{2}$.  Therefore, $\lambda\gtrsim 2\chi^{-1/2}\delta$, which sets lower limit on the filament radius to be about twice the plasma skin depth.  

On the other hand, the current inside the filament $|I|\sim\lambda^2\chi n e c$ cannot exceed the Alfv\'en limit $I_{\rm A}\sim A \gamma m c^3/e$, where $A\sim 1$ is a numerical form factor. (In principle, the current can exceed the limit if the perpendicular velocities in the filament greatly exceed the parallel velocities, which is not the case here.)  
From this one can conclude that $\lambda\lesssim 2\pi^{1/2}\chi^{-1/2}\delta$.  Evidently, the limit is saturated when charge separation is almost complete in a filament with radius just slightly larger than the plasma skin depth. 
The above two constraints pen the radius of the filament to be just slightly larger than  the plasma skin depth, $\lambda\sim 2\delta$, where we have taken into account that practically all particles become confined within filaments, $\chi\sim 1$, at the end of the charge separation phase.

Similarly, for the magnetic field $|B|\sim 2|I|/\lambda c$ to be dynamically important, the ratio of the magnetic to mechanical and thermal energy density $\epsilon_B\equiv B^2/8\pi\gamma mc^2n$ should be of the order of unity.  This ratio can be expressed as $\epsilon_B\sim \onehalf|I/I_{\rm A}|^2(\lambda/\delta)^{-2}$.  Since $|I|\leq I_{\rm A}$, it follows that $\lambda\gg\delta$ would imply $\epsilon_B\ll 1$, rendering the magnetic field dynamically unimportant.  Indeed, $\epsilon_B\sim 0.1-1$ has been observed in PIC simulations of shell collisions in pair plasmas (e.g.,~\citealt{Kazimura:98,Gruzinov:01,Silva:03,Medvedev:05}).

\section{The Precursor}
\label{sec:precursor}

If the counterstreaming particles at $x_0$ are thermal with a Maxwellian tail extending to energies $\gamma'mc^2> q'\phi(x_0)$, a steady stream of particles escape into the upstream.  These particles run ahead of the shock wave as a ``precursor.'' In principle, the precursor can initiate transverse Weibel instability and thus transfer energy ahead of the shock. This would  invalidate the premise that the upstream is cold and field-free as it enters the charge separation layer.  Fortunately, finite perpendicular motions of the counterstream particles in the upstream frame can suppress the instability if the density of the escaping particles is not too large.  We here discuss conditions for the stability of the precursor.

The approximate, waterbag-type dispersion relation for transverse electromagnetic modes, assuming that the upstream is cold and the precursor is warm, is derived in Appendix \ref{sec:dispersion_relation}.  To check whether the upstream is stable in the presence of a low density precursor, we consider the case $N'_{\rm free}\ll \gamma_{-\infty}^{-1}$ where $N'_{\rm free}\equiv n'/n_{-\infty}$ is the ratio of the density of the escaping precursor to the density of the upstream in the frame of the shock. With this assumption, the upstream is nonrelativistic in the special frame in which the perpendicular electric field vanishes (see Appendix \ref{sec:dispersion_relation}). The unstable wave numbers must satisfy equation (\ref{eq:nonrelativistic_unstable_modes}) and thus instability takes place when (an equivalent relation was also derived in \citealt{Silva:02})
\beq
\label{eq:condition_unstable_precursor}
\frac{n'/\gamma'}{n_{-\infty}/\gamma_{-\infty}}\gtrsim {\beta_\perp'}^2 ,
\eeq
where $\beta_\perp'$ is the typical perpendicular velocity of the precursor particles in the frame of the current defined in \S~\ref{sec:ballistics}, which is related to the same quantity in the shock frame via $\beta_\perp'=\beta_r'/\tilde\gamma(1+\beta_r'\tilde\beta)\lesssim 1/\gamma_{-\infty}$.  Given that majority of the precursor particles have $\gamma'\sim\gamma_{-\infty}$, the upstream-precursor system is stable as long as $N'_{\rm free}\lesssim \gamma_{-\infty}^{-2}$ and is unstable for $N'_{\rm free}\gtrsim \gamma_{-\infty}^{-2}$.  Therefore to check whether the precursor initiates transverse Weibel instability ahead of the shock, we must compare $N'_{\rm free}$ with $\gamma_{-\infty}^{-2}$. 

The requirement that the precursor be stable to the development of the transverse Weibel instability can be seen as setting a condition on the number of particles that are contained in the unconfined, high-energy Maxwellian tail.  Within the model presented in \S~\ref{sec:counterstream}, the density of these particles is 
\bea
\label{eq:n_free}
N'_{\rm free}&\sim& N'(x_0) e^{q'\phi(x_0)/\gamma_0mc^2}\nonumber\\
&\sim&  N'(x_0) e^{-L^2\gamma_{-\infty}/4\gamma_0} .
\eea
In an ultrarelativistic shock, energy conservation dictates that $\gamma_0\sim\gamma_{-\infty}/3\sqrt{2}$.  We have argued in \S~\ref{sec:diameter} that $L\sim2$. Then from equation (\ref{eq:n_free}), we have that $N'_{\rm free}/N'(x_0) \sim e^{-3\sqrt{2}}\approx 0.014$. The precursor is stable if $N'_{\rm free}< \gamma_{-\infty}^2$, or $\gamma_{-\infty}^2N'(x_0)<e^{3\sqrt{2}}\approx 70$.  Thus if $\gamma_{-\infty} > e^{3/\sqrt{2}}\approx 8$, the precursor can be stable only if $N'< 1$, which means that the density of the counterstream particles at $x_0$ is less than the density of the upstream.  However the density of the counterstreaming particles must become comparable to that of the upstream at some point in the transition layer, since one-third of the downstream particles have negative velocities in the frame of the shock. Therefore in shocks with large Lorentz factors, compression and isotropization must take place downstream of the charge separation layer.

We speculate that if the upstream medium is weakly magnetized, reflection of precursor particles back toward the shock might play a role in the injection of particles into the diffusive accelerator mechanism of the shock front.

\section{Discussion}
\label{sec:discussion}

In the hydrodynamic limit, the flow crossing an adiabatic relativistic shock wave is compressed by the factor $n_{+\infty}/n_{-\infty}\sim 3$.  We saw that the structure of the transition layer is sensitive to the strength of the electrostatic potential and the width of the current filaments.  If the potential is strong enough to decelerate the incoming upstream to a nonrelativistic mean bulk velocity in the frame of the shock (note that each particle is still relativistic), compression can take place in the electrostatic charge separation layer, although, of course, thermalization still requires a dynamical electromagnetic field. If the flow remains relativistic through the charge separation layer, the shock compression must take place afterwards, and the charge separation layer is followed by a compression layer.  We elaborate briefly on the latter possibility.

In the charge separation layer, current filaments do not interact because the perpendicular displacement of the upstream particles is of the order of the filament size.  Also, there is a near-perfect balance of the magnetic field (which attracts like currents) and the electric field (which repels them).  Therefore, the charge separation layer is approximately two-dimensional.  This two-dimensional picture breaks down when the upstream particles have crossed the filaments once and charge separation has reached maximum.  At that point, the filaments start to interact and the picture becomes three-dimensional.   These three-dimensional processes also govern the decay of the magnetic field.

Magnetic interactions cause rearrangement among the filaments (unfolding, mergers, etc.). Even noninteracting filaments do not remain parallel to the direction of propagation of the shock.  Based on an analogy with magnetohydrodynamical instabilities in magnetically-confined columns, initially straight collisionless current filaments should be unstable to a kink or pinch-like mode.  The filaments bend, buckle, and fold.  Time-varying electric fields associated with the rearrangement scatter particles inelastically to large pitch angles, thereby aiding diffusion of the momentum, rendering the flow isotropic and eventually thermalized in the frame of the downstream plasma. Isotropization in the downstream frame implies compression.

The results thus derived apply to $e^-e^+$ shocks.  Does the same picture describe electron-ion shocks?  Simulations of colliding electron-ion shells \citep{Frederiksen:04} reveal that the electrons thermalize before the ions do.  If the transition layer contains ion filaments, as the simulations suggest to be the case, the electrons are scattered by the electromagnetic field of the ion filaments.  Charge separation among the ions takes place in the background of hot electrons.  The mechanism of charge separation presented in \S~\ref{sec:separation} and \S~\ref{sec:counterstream} could apply to the ustream ions.  They pinch into filaments as a small amount of hot electrons are trapped within each upstream ion   filament and cancel some of the ionic self-electric field.  The upstream ion filaments are much like the $e^-e^+$ filaments because the downstream electrons moving toward the upstream have comparable energies, and thus comparable inertia, as the downstream ions.  At this point we cannot explain the force that keeps the downstream ions from escaping into the upstream, although we expect that an electrostatic field associated with the upstream electrons is involved.

\section{Conclusions}
\label{sec:conclusions}

We here summarize the main conclusions of the paper:

1. The first generation of magnetically-confined current filaments that arise in the saturated state of the transverse Weibel instability in collisionless shocks are nonneutral and nonuniform.  

2. Two interpenetrating components in the shock transition layer are the upstream particles and the downstream particles moving toward the upstream in the frame of the shock (``the counterstream''). Both components  are charge-separated.

3. Space charge in the filaments gives rise to an electrostatic potential; we call the leading, electrostatic edge of the transition layer ``the charge separation layer.''

4. In the charge separation layer the charge density and the magnetic energy density increase in the parallel direction.

5. The electrostatic potential confines most of the counterstream particles to the charge separation layer, thereby preventing their escape into the upstream.  

6. The radii of first-generation filaments must be just slightly larger than the plasma skin depth.

7. A small fraction of downstream particles, ``the precursor,'' escape into the upstream.  These particles may be important for diffusive acceleration in the shock.

8. The precursor moving through the upstream must be stable to the development of the transverse Weibel instability, which sets a condition on the density of counterstreaming particles in the charge separation layer.

9. Shock compression is completed downstream of the charge separation layer, rather than in the layer itself.

10. Downstream of the charge separation layer filaments merge on a dynamical time scale and the electromagnetic field is dynamic, facilitating isotropization and thermalization of the plasma.

\acknowledgements

It is a pleasure to thank J.~Arons, S.~Cowley, P.~Goldreich, A.~K\"onigl, \AA.~Nordlund, T.~Piran, and R.~Sari for stimulating discussions.  M.~M.\ and E.~N.\ were supported at Caltech, respectively, by a postdoctoral fellowship and a senior research fellowship from the Sherman Fairchild Foundation.  A.~S.\ acknowledges support provided by NASA through Chandra Fellowship
grant PF2-30025 awarded by the Chandra X-Ray Center, which is operated
by the Smithsonian Astrophysical Observatory for NASA under contract
NAS8-39073.

\appendix

\section{Dielectric Permitivity Tensor}

We work in the orthonormal basis $(\hat {\bf e}_\parallel,\hat {\bf e}_{\perp1},\hat {\bf e}_{\perp2})$ where $\hat {\bf e}_\parallel$ is parallel to the direction of propagation of the shock.  Consider linear electromagnetic perturbations of the field-free plasma of the form
\bea
{\bf E}({\bf x},t)&=&{\bf E}({\bf k},\omega) \exp[i({\bf k}\cdot {\bf x}-\omega t)] , \nonumber\\ 
{\bf B}({\bf x},t)&=&{\bf B}({\bf k},\omega) \exp[i({\bf k}\cdot {\bf x}-\omega t)] ,
\eea
subject to the condition ${\bf k}\times {\bf E}({\bf k},\omega)=(\omega/c){\bf B}({\bf k},\omega)$.  To study filamentation in the transition layer of the shock, we consider transverse modes, $k_\parallel=0$.  This is a Lorentz-invariant choice as long as $\omega$ is purely imaginary.  Components of the dielectric permitivity tensor $\varepsilon$ can be evaluated using standard methods (e.g., \citealt{Krall:73}).  The perturbed electric field satisfies the linearized Maxwell-Vlasov equations ${\bf D}\cdot {\bf E}=0$ where ${\bf E}=(E_\parallel,E_{\perp1},E_{\perp2})$ and $D_{ij}=\varepsilon_{ij}+(c^2/\omega^2)(k_ik_j-k^2\delta_{ij})$.  Note that $\varepsilon$ and ${\bf E}$ refer to the permitivity tensor and the electric field in Fourier space; explicit dependence on the wave vector is omitted here.  We use the index $\alpha$ to label distinct species of particles with phase space distribution functions $f_\alpha({\bf p})$, particle rest masses $m_\alpha$, concentrations in the observer frame $n_\alpha$, and plasma frequencies $\omega_{{\rm p}\alpha}\equiv(4\pi n_\alpha e^2/m_\alpha)^{1/2}$.  Components of the permitivity tensor then equal
\bea
\label{eq:permitivity}
\varepsilon_{ij}= \delta_{ij}+ \sum_\alpha \frac{\omega_{{\rm p}\alpha}^2}{\omega^2}  m_\alpha\int
\left(
\frac{\partial f_\alpha}{\partial p_{j}}+
 v_{j}\frac{
{\bf k}\cdot\nabla_{{\bf p}} f_\alpha  }
{\omega-{\bf k}\cdot {\bf v}}
\right)
v_{i} d{\bf p} ,
\eea
where the velocities are related to the momenta via ${\bf p}=\gamma m_\alpha {\bf v}$.

\section{Dispersion Relation}
\label{sec:dispersion_relation}

We choose to work with the waterbag distribution function for each plasma component
\beq
\label{eq:waterbag}
f_\alpha({\bf p}) = \frac{1}{\pi {p_{\rm \perp\alpha}}^2} \delta(p_\parallel-p_{\parallel\alpha})
H(p_{\rm \perp\alpha}^2-p_\perp^2) ,
\eeq
where $p_{\parallel\alpha}$ is the bulk momentum of the plasma, $p_{\rm \perp\alpha}$ is the maximum perpendicular momentum, $\delta(x)$ is the Dirac $\delta$-function, and $H(x)$ is the Heaviside step function.  The distribution function in equation (\ref{eq:waterbag}) assumes that the plasma is cold in the parallel direction. Note that the distribution is isotropic around the axis of propagation of the shock. The waterbag-type distribution functions in \citet{Medvedev:99}, \citet{Silva:02}, \citet{Fonseca:03}, and \citet{Wiersma:04} are not isotropic in this way.  The differences stemming from the particular choice of distribution function, however, do not seem to affect the instability growth rates substantially.

Current filaments parallel to direction of propagation of the shock form from electromagnetic modes with purely perpendicular magnetic field, $B_\parallel=0$, and purely parallel electric field, ${\bf E}_\perp=0$.  The latter condition is violated in a general reference frame.  There exists a unique Lorentz frame in which the parallel and perpendicular electric fields decouple and the linearized Maxwell-Vlasov equation is block-diagonal, that is, the cross term vanishes $D_{\parallel\perp i}=0$.  In this frame, the dispersion relation for the linear development of the filamentation instability is simply $D_{\parallel\parallel}=0$. Substituting equation (\ref{eq:waterbag}) into the first and second of equations (\ref{eq:permitivity}) yields
\bea
\label{eq:d_general}
D_{\parallel\parallel}&=&1-\frac{c^2k^2}{\omega^2}
+\sum_\alpha \frac{\omega_{{\rm p}\alpha}^2 /\gamma_\alpha}{\omega^2} 
\left[- \frac{2(1+\gamma_\alpha\gamma_{\parallel\alpha} -\gamma_{\parallel\alpha}^2)}
{\gamma_{\parallel\alpha}(\gamma_\alpha+\gamma_{\parallel\alpha})}
+\beta_{\parallel\alpha}^2\frac{2}{\beta_{\perp\alpha}^2}\left(1-\frac{\omega}
{\sqrt{\omega^2-c^2k^2\beta_{\perp\alpha}^2}}\right) \right] ,\nonumber\\
D_{\parallel\perp i}&=& \frac{\omega}{ck} \sum_\alpha \frac{\omega_{{\rm p}\alpha}^2/\gamma_\alpha}{\omega^2}\beta_{\parallel\alpha}  \frac{2}{ \beta_{\perp\alpha}^2} \left(1-\frac{\omega}{\sqrt{\omega^2-c^2k^2\beta_{\perp\alpha}^2}}\right) ,
\eea
where, omitting subscript $\alpha$, we used the definitions $\beta_{\parallel}=p_{\parallel}/\gamma m c$, $\beta_{\perp}=p_{\perp}/\gamma m c$, $\beta^2 = \beta_{\parallel}^2+\beta_{\perp}^2$, $\gamma=(1-\beta^2)^{-1/2}$, and $\gamma_\parallel=(1+\gamma^2\beta_{\parallel}^2)^{1/2}$.  

Consider large phase velocity perturbations $\omega/k \gg c \beta_{\perp}$ in cold plasmas $\gamma\beta_\perp\ll 1$.  Note that
\beq
\label{eq:expand_gamma_term}
\frac{2(1+\gamma \gamma_\parallel -\gamma_\parallel^2)}{\gamma_{\parallel}(\gamma+\gamma_{\parallel})}=\frac{1}{\gamma^2}\left\{1+\left(\frac{3}{4}+\frac{\gamma^2}{2}\right)\beta_\perp^2+{\cal O}[(\gamma\beta_\perp)^4]\right\}
\eeq
and 
\bea
\label{eq:expand_omega_term}
\frac{2}{\beta_\perp^2}\left(1-\frac{\omega}
{\sqrt{\omega^2-c^2k^2\beta_{\perp}^2}}\right)&=&-\left(\frac{ck}{\omega}\right)^2
-\frac{3}{4}\left(\frac{ck}{\omega}\right)^4\beta_\perp^2+{\cal O}[(ck\beta_\perp/\omega)^4] \nonumber\\
&=&  -\frac{c^2k^2}{\omega^2-\threequarters c^2k^2{\beta_\perp'}^2}+{\cal O}[(ck\beta_\perp/\omega)^4].
\eea
Thus, to the lowest order in $\beta_{\perp\alpha}$, the condition $D_{\parallel\perp i}=0$, required for $E_\perp=0$ and $B_\parallel=0$, reduces to
\beq
\label{eq:center}
\sum_\alpha \frac{\omega_{{\rm p}\alpha}^2}{\gamma_\alpha}\beta_{\parallel\alpha}\propto \sum_\alpha \frac{n_{\alpha0}}{m_\alpha}\beta_{\parallel\alpha} =0 ,
\eeq
where we have defined rest-frame particle concentrations $n_{\alpha0}\equiv n_\alpha/\gamma_\alpha$. The frame in which the perpendicular electric field vanishes is not the rest frame of the plasma.

\section{Stability}

Consider a plasma containing a dense, infinitely cold component, and a counterstreaming low-density component with finite perpendicular temperature.  Following the notation of \citet{Akhiezer:75}, we add a prime to the quantities pertaining to the low-density component and dispense with the subscript $\alpha$.  Assume that $n/\gamma\gg n'/\gamma'$. Then the low-density component is ultrarelativistic, $\gamma'\gg1$ and $\beta_\parallel'\sim-1$.  Equation (\ref{eq:center}) then implies that $\omega_{\rm p}^2\beta_\parallel/\gamma\sim{{\omega}_{\rm p}'}^{2}/\gamma'$. Defining the ratio of the rest-frame concentrations $\alpha\equiv n_0'/n_0\ll1$, we have that $\beta_\parallel\sim\alpha\ll1$ and $\gamma\sim 1$. In the same context, \citet{Silva:02} define the concentration ratio in the frame in which the denser component is at rest.  \

In the frame with a vanishing perpendicular electric field, the dispersion relation simplified using (\ref{eq:expand_omega_term}) and (\ref{eq:expand_gamma_term}) reads
\beq
1-\left(\frac{ck}{\omega}\right)^2-\frac{\omega_{\rm p}^2}{\omega^2}
\left[1
+\beta_{\parallel}^2\left(\frac{ck}{\omega}\right)^2\right] -
\frac{{\omega_{\rm p}'}^2/\gamma'}{\omega^2}
\left(\frac{1}{{\gamma'}^2}+
\frac{c^2k^2}{\omega^2-\threequarters c^2k^2{\beta_\perp'}^2}
\right) = 0.
\eeq
or, using ${\omega_{\rm p}'}^2/\gamma'\sim\alpha \omega_{\rm p}^2$ and keeping the lowest order terms
\beq
1-\left(\frac{ck}{\omega}\right)^2-\frac{\omega_{\rm p}^2}{\omega^2}
-\alpha
\frac{{\omega_{\rm p}}^2}{\omega^2}
\frac{c^2k^2}{(\omega^2-\threequarters c^2k^2{\beta_\perp'}^2)} = 0.
\eeq
This equation is stable for all values of the wave vector when $4\alpha<3{\beta_\perp'}^2$.  For  $4\alpha>3{\beta_\perp'}^2$, the condition for instability is
\beq
\label{eq:nonrelativistic_unstable_modes}
k^2 \lesssim \frac{\omega_{\rm p}^2}{c^2}\left(\frac{4}{3}\frac{\alpha}{{\beta_\perp'}^2}-1\right) .
\eeq
The growth rate and the wave number of the fastest growing mode are
\beq
\omega_{\rm max}^2\approx -\omega_{\rm p}^2 \left(\sqrt{\alpha}-\sqrt{\frac{3}{4}}\beta_\perp'\right)^2 , \ \ \ \  k^2_{\rm max}\approx\frac{\omega_{\rm p}^2}{c^2}\left(\sqrt{\frac{4}{3}}\frac{\sqrt{\alpha}}{{\beta_\perp'}}-1\right) .
\eeq

Next, consider the case in which both components of the plasma are relativistic, have similar densities, and one component is infinitely cold while the other has finite perpendicular temperature.  Here $\beta_\parallel\sim\beta_\parallel'\sim 1$ and thus $\omega_{\rm p}^2/\gamma \sim {\omega_{\rm p}'}^2/\gamma'$.  Relevant terms in the dispersion relation then read 
\beq
\label{eq:dispersion_similar_density}
1-\left(\frac{ck}{\omega}\right)^2-2\frac{\omega_{\rm p}^2/\gamma}{\omega^2}
\frac{c^2k^2}{(\omega^2-\frac{3}{8} c^2k^2 {\beta_\perp'}^2)}=0 .
\eeq
The condition for instability is then
\beq
k^2\lesssim \frac{16}{3} \frac{\omega_{\rm p}^2}{\gamma c^2{\beta_\perp'}^2} .
\eeq
The growth rate and the wave number of the fastest growing mode are 
\beq
\omega_{\rm max}^2 \approx -2\frac{\omega_{\rm p}^2}{\gamma}\left(1-\sqrt{\frac{3}{2}}\beta_\perp'\right) , \ \ \ \ k_{\rm max}^2\approx 4\frac{\omega_{\rm p}^2}{\gamma c^2}\left(\sqrt{\frac{2}{3}}\frac{1}{\beta_\perp'}-1\right) .
\eeq
Note that here $\omega_{\rm p}$ and $\gamma$ are the plasma frequency and Lorentz factor of the infinitely cold component.  We made no assumption about which component is denser, but have assumed that the densities are similar enough for both components to be relativistic in the frame in which the perpendicular electric field vanishes.


\begin{thebibliography}{99}

\bibitem[Akhiezer et al.(1975)]{Akhiezer:75} Akhiezer, A.~I., Akhiezer, I.~A., Polovin, R.~V., Sitenko, A.~G., \& Stepanov, K.~N.\ 1975, Plasma Electrodynamics (Pergamon: Oxford)

\bibitem[Alfv{\' e}n(1939)]{Alfven:39} Alfv{\' e}n, H.\ 1939, Phys. Rev., 55, 425

\bibitem[Blandford \& McKee(1976)]{Blandford:76} Blandford, R.~D., \& McKee, C.~F.\ 1976, Phys. Fluids, 19, 1130

\bibitem[Bret, Firpo, \& Deutsch(2004)]{Bret:04} Bret, A., Firpo, M.~C., \& Deutsch, C.\ 2004, \pre, 70, 046401 

\bibitem[Bret, Firpo, \& Deutsch(2005)]{Bret:05} Bret, A., Firpo, M.~C., \& Deutsch, C.\ 2005, \prl, 94, 115002

\bibitem[Davidson(1974)]{Davidson:74} Davidson, R.~C.\ 1974, Frontiers in Physics, 43 (Reading: W.~A.~Benjamin)



\bibitem[Fonseca et al.(2003)]{Fonseca:03} Fonseca, R.~A., Silva, L.~O., Tonge, J.~W., Mori, W.~B., \& Dawson, J.~M.\ 2003, Phys. Plasmas, 10, 1979

\bibitem[Frederiksen et al.(2004)]{Frederiksen:04} Frederiksen, J.~T., Hededal, C.~B., Haugb{\o}lle, T., Nordlund, {\AA}.\ 2004, \apjl, 608, L13

\bibitem[Fried(1959)]{Fried:59} Fried, B.~D.\ 1959, Phys. Fluids, 2, 337

\bibitem[Gruzinov(2001)]{Gruzinov:01} Gruzinov, A.\ 2001, \apjl, 563, L15

\bibitem[Honda(2000)]{Honda:00} Honda, M.\ 2000, Phys. Plasmas, 7, 1606

\bibitem[Jaroschek, Lesch, \& Treumann(2004)]{Jaroschek:04} Jaroschek, C.~H., Lesch, H., \& Treumann, R.~A.\ 2004, \apj, 616, 1065

\bibitem[Jaroschek, Lesch, \& Treumann(2005)]{Jaroschek:05} Jaroschek, C.~H., Lesch, H., \& Treumann, R.~A.\ 2005, \apj, 618, 822

\bibitem[Kato(2005)]{Kato:05} Kato, T.~N.\ 2005, preprint (physics/0501110)

\bibitem[Kazimura et al.(1998)]{Kazimura:98} Kazimura, Y., Sakai, J.~I., Neubert, T., \& Bulanov, S.~V.\ 1998, \apjl, 498, L183

\bibitem[Krall \& Trivelpiece(1973)]{Krall:73} Krall, N.~A., \& Trivelpiece, A.~W.\ 1973, Principles of Plasma Physics (McGraw-Hill: New York)

\bibitem[Lee \& Lampe(1973)]{Lee:73} Lee, R., \& Lampe, M.\ 1973, Phys. Rev. Lett., 31, 1390


\bibitem[Medvedev \& Loeb(1999)]{Medvedev:99} Medvedev, M.~V., \& Loeb, A.\ 1999, \apj, 526, 697

\bibitem[Medvedev et al.(2005)]{Medvedev:05} Medvedev, M.~V., Fiore, M., Fonseca, R.~A., Silva, L.~O., \& Mori, W.~B.\ 2005, \apjl, 618, L75

\bibitem[Moiseev \& Sagdeev(1963)]{Moiseev:63} Moiseev, S.~S., \& Sagdeev, R.~Z.\ 1963, J. Nuclear Energy, 5, 43

\bibitem[Montgomery \& Joyce(1969)]{Montgomery:69} Montgomery, D.~C., \& Joyce, G.\ 1969, J. Plasma Phys., 4, 1

\bibitem[Nishikawa et al.(2005)]{Nishikawa:05} Nishikawa, K.~I., Hardee, P., Richardson, G., Preece, R., Sol, H., \& Fishman, G.~J.\ 2005, \apj, 622, 927

\bibitem[Raadu(1989)]{Raadu:89} Raadu, M.~A.\ 1989, \physrep, 178, 25

\bibitem[Silva et al.(2002)]{Silva:02} Silva, L.~O., Fonseca, R.~A., Tonge, J.~W., Mori, W.~B., \& Dawson, J.~M.\ 2002, Phys. Plasmas, 9, 2458

\bibitem[Silva et al.(2003)]{Silva:03} Silva, L.~O., Fonseca, R.~A., Tonge, J.~W., Dawson, J.~M., Mori, W.~B., \& Medvedev, M.~V.\ 2003, \apjl, 596, L121


\bibitem[Weibel(1959)]{Weibel:59} Weibel, E.~S.\ 1959, Phys. Rev. Lett., 2, 83

\bibitem[Wiersma \& Achterberg(2004)]{Wiersma:04} Wiersma, J., \& Achterberg, A.\ 2004, \aap, 428, 365

\bibitem[Yoon \& Davidson(1987)]{Yoon:87} Yoon, P.~H., \& Davidson, R.~C.\ 1987, \pra, 35, 2718

\end{thebibliography}
\end{document}